\documentclass[aps,nofootinbib,prl,groupedadthe $k$-gap closes anddress,twocolumn]{revtex4}
\pdfoutput=1
\usepackage{graphicx}
\usepackage{amsmath,amsfonts}
\usepackage{gensymb}
\usepackage{physics}
\usepackage{color,ulem}

\usepackage{hyperref}
\begin{document}

\title{Maxwell interpolation and close similarities between liquids and holographic models}
\author{M. Baggioli$^{1,2}$}
\email{matteo.baggioli@uam.es}
\author{K. Trachenko$^{3}$}
\email{k.trachenko@qmul.ac.uk}
\address{$1.$ Crete Center for Theoretical Physics, Institute for Theoretical and Computational Physics, Department of Physics, University of Crete, 71003 Heraklion, Greece}
\address{$2.$ Instituto de Fisica Teorica UAM/CSIC, c/Nicolas Cabrera 13-15,
Universidad Autonoma de Madrid, Cantoblanco, 28049 Madrid, Spain.}
\address{$3.$ School of Physics and Astronomy, Queen Mary University of London, Mile End Road, London, E1 4NS, UK}

\begin{abstract}
We show that liquids and certain holographic models are strikingly similar in terms of several detailed and specific properties related to their energy spectra. We consider two different holographic models and ascertain their similarity with liquids on the basis of emergence of the gap in transverse momentum space and the functional form of the dispersion relation. Furthermore, we find that the gap increases with temperature, the relaxation time governing the gap decreases with temperature and, finally, the gap is inversely proportional to the relaxation time as in liquids. On this basis, we propose that the general idea involved in Maxwell-Frenkel approach to liquids can be used to understand holographic models and their strongly-coupled field theory counterparts in a non-perturbative way.
\end{abstract}


\maketitle

\section{Introduction}
Many interesting effects in quantum field theory (QFT) are related to strongly-coupled dynamics. These problems can not be solved by the perturbative approaches commonly used in QFT. However, the proposal of the correspondence between QFT and gravity models (AdS-CFT corrrespondence) has opened the way to approach strongly-coupled QFT problems by addressing the corresponding weakly-interacting gravitational duals \cite{ads1,ads2}.

For the same reason of strong coupling, a theory of liquids was believed to be impossible to construct in a general form \cite{landau}. Perturbation theories do not apply to liquids because the inter-atomic interactions are strong. Solid-based approaches seemingly do not apply to liquids either: its unclear how to apply the traditional harmonic expansion around the equilibrium positions because the equilibrium lattice does not exist due to particle re-arrangements that enable liquids to flow. This combination of strong interactions and large particle displacements has proved to be the ultimate problem in understanding liquids theoretically, and is known as the ``absence of a small parameter''.

The absence of traditional simplifying features in the liquid description does not mean that the problem can not be solved in some other way, including attempting the first-principles approach using the equations of motion. However, this involves solving a large number of non-linear equations, an exponentially complex problem not currently tractable \cite{ropp}.

In this paper, we find striking similarities between liquids and certain holographic systems. In particular we will underpin the common features on the basis of Maxwell interpolation giving rise to a specific dispersion relation of the type:
\begin{equation}
\omega^2\,\approx\,k^2\,-\,k_g^2
\end{equation}
which we call the {\it k-gapped} dispersion relation.

In liquids, the $k$-gap gives the upper cutoff of wavelengths at which the shear waves can propagate and is related to liquid relaxation time representing the average time between molecular rearrangements \cite{frenkel}. In Maxwell viscoelastic model discussed below, the relaxation time can be written as:
\begin{equation}
\tau_M\,=\,\frac{\eta}{G_{\infty}}\label{tmax}
\end{equation}
where $\eta$ is the shear viscosity and $G_{\infty}$ the instantaneous shear modulus. At the microscopic level, Frenkel's theory \cite{frenkel} has identified the relaxation time with the average time between consecutive molecular rearrangements, the picture that has become widely accepted since \cite{dyre}.

In holographic models, the relaxation time arises from the intrinsic dissipative dynamics connected to the finite temperature setups and in particular to the dissipative nature of the black hole event horizon. Although it may be hard to rigorously prove that this mechanism is the same as in non-relativistic liquids, we will see that the commonalities are intriguing.
First, the dispersion relation found in the holographic models is the same as predicted by Maxwell interpolation. Second, the $k$-gap increases with temperature as in liquids and the corresponding relaxation time shows a temperature behaviour similar to the Vogel-Tammann-Fulcher law seen in liquids. The strong similarities suggest an underlying and fundamental principle behind the physics of the $k$-gap which has not yet been identified.

In the first section \ref{sect1}, we review the theory of phonons in liquids, Maxwell-Frenkel approach and the emergence of the $k$-gap. In the second section \ref{sec2}, we present two simple holographic models which display the $k$-gap and show other interesting similarities with liquids. In the final section \ref{sec3}, we conclude and discuss possible future directions.

\section{Maxwell-Frenkel approach to liquids and the emergence of $k$-gap}\label{sect1}
Recent progress in understanding liquid thermodynamics followed from considering what kind of collective modes (phonons) can propagate in liquids and supercritical fluids \cite{ropp}. It has been ascertained that solid-like transverse modes can propagate in liquids but, interestingly, they develop a gap in $k$, or {\it momentum} space, with the gap growing with the inverse of liquids relaxation time \cite{prl}. This enabled us to discuss and understand liquid thermodynamics on the basis of phonons, as is done in the solid-state theory.

We start with recalling how liquid transverse modes develop a gap in momentum space. This programme starts with Maxwell interpolation:

\begin{equation}
\frac{d\mathfrak{S}}{dt}=\frac{\mathfrak{P}}{\eta}+\frac{1}{G}\frac{d\mathfrak{P}}{dt}
\label{a1}
\end{equation}
\noindent where $\mathfrak{S}$ is shear strain, $\eta$ is viscosity, $G$ is shear modulus and $\mathfrak{P}$ is shear stress.

Eq. (\ref{a1}) reflects the Maxwell's proposal \cite{maxwell} that the shear response in a liquid is the sum of viscous and elastic responses given by the first and second right-hand side terms. Notably, neither elastic nor the dissipative term containing the viscosity are introduced as a small perturbation: both elastic and viscous deformations are treated in (\ref{a1}) on equal footing. This implies that hydrodynamics and elasticity can be equally good starting point of liquid description. We will return to this point below.

Frenkel proposed \cite{frenkel} to represent the Maxwell interpolation by introducing the operator $A$ as $A=1+\tau\frac{d}{dt}$ so that Eq. (\ref{a1}) can be written as $\frac{d\mathfrak{S}}{dt}=\frac{1}{\eta}A \mathfrak{P}$. Here, $\tau$ is the Maxwell relaxation time $\frac{\eta}{G}$. Frenkel's idea was to generalize $\eta$ to account for liquid's short-time elasticity as $\frac{1}{\eta}\rightarrow\frac{1}{\eta}\left(1+\tau\frac{d}{dt}\right)$ and use this $\eta$ in the Navier-Stokes equation as $\nabla^2{\bf v}=\frac{1}{\eta}\rho\frac{d{\bf v}}{dt}$, where ${\bf v}$ is velocity, $\rho$ is density and the full derivative is $\frac{d}{dt}=\frac{\partial}{\partial t}+{\bf v\nabla}$. We have carried this idea forward \cite{ropp} and, considering small ${\bf v}$, wrote

\begin{equation}
c^2\frac{\partial^2v}{\partial x^2}=\frac{\partial^2v}{\partial t^2}+\frac{1}{\tau}\frac{\partial v}{\partial t}
\label{gener3}
\end{equation}

\noindent where $v$ is the velocity component perpendicular to the $x$ direction, $\eta=G\tau=\rho c^2\tau$ and $c$ is the shear wave velocity.

Eq. (\ref{gener3}) can also be obtained by starting with the solid-like elastic equation for the non-decaying wave and, using Maxwell interpolation (\ref{a1}), generalizing the shear modulus to include the viscous response \cite{pre}. This shows that the hydrodynamic approach commonly applied to liquids \cite{hydro} is not a unique starting point and that the solid-like elastic approach is equally legitimate, implying an interesting symmetry of the liquid description. This is consistent with elastic and viscous terms being treated on equal footing in (\ref{a1}) as mentioned above.

\begin{figure}
\includegraphics[width=0.7\linewidth]{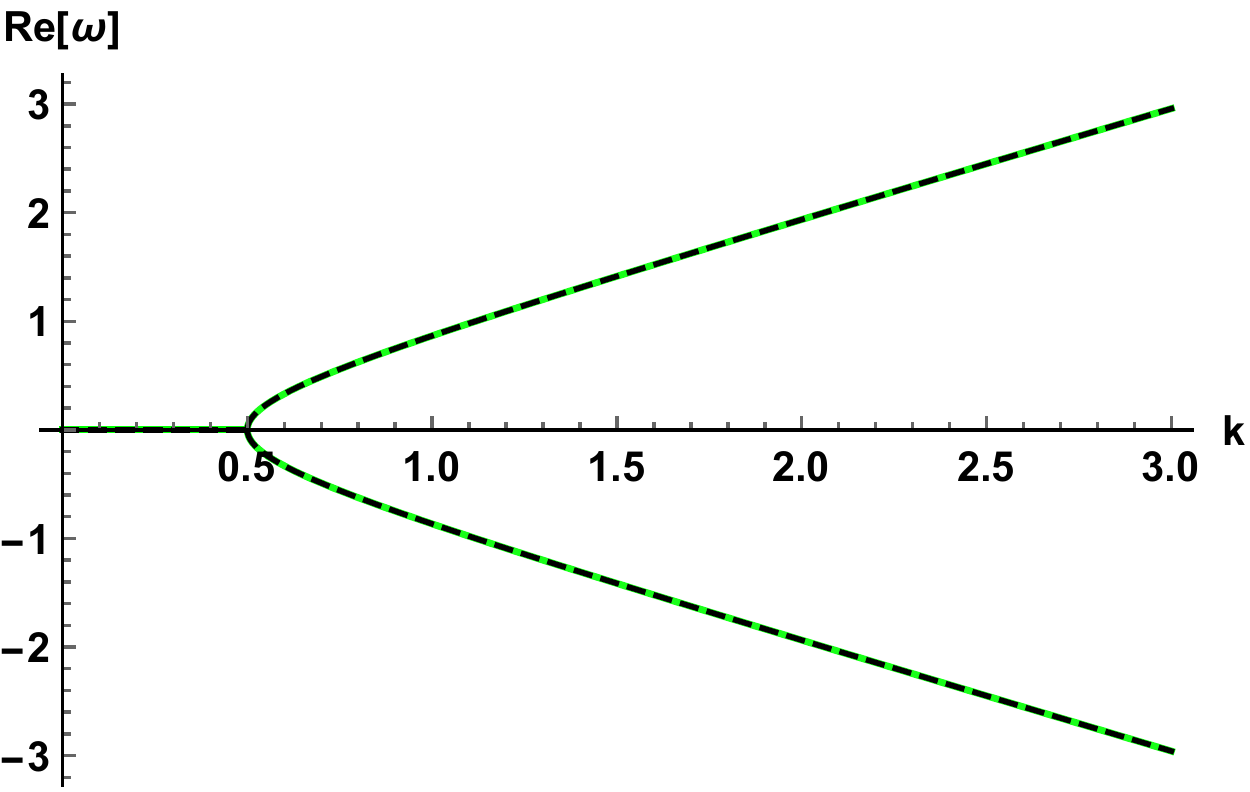}
\vspace{0.2cm}
\includegraphics[width=0.7\linewidth]{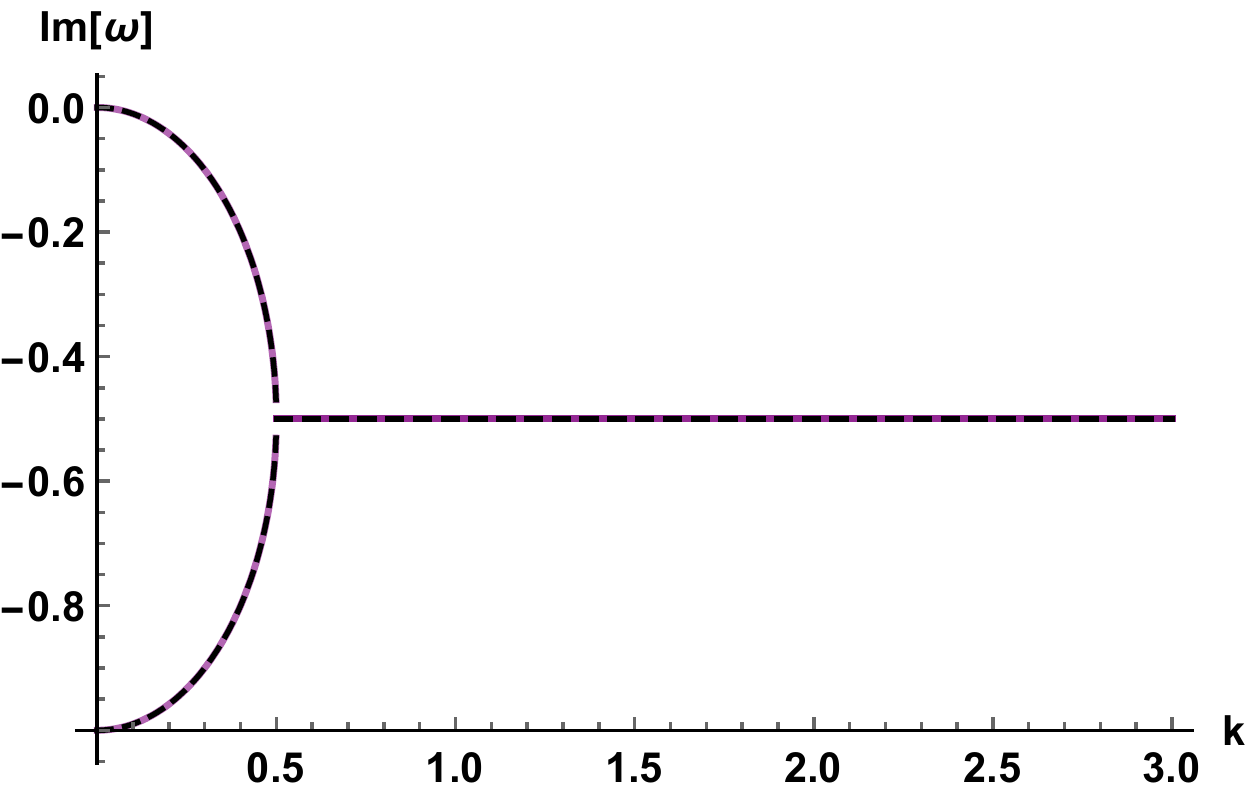}
\caption{The dispersion relation of the two modes in Eq. \eqref{omega}. At $k=k_{gap}=0.5$ the collision happens and a propagating mode with positive real part appears. The parameters are fixed to $c=\tau=1$.}
\label{fig0}
\end{figure}

In contrast to the Navier-Stokes equation, Eq. (\ref{gener3}) contains the second time derivative of $v$ and hence allows for propagating waves. We solved Eq. (\ref{gener3}) \cite{ropp} by seeking a plane-wave solution as $v=v_0\exp\left(i(kx-\omega t)\right)$. This gives
\begin{equation}
\omega^2+\omega\frac{i}{\tau}-c^2k^2=0
\label{quad}
\end{equation}
\noindent and the following dispersion relation
\noindent
\begin{equation}
\omega=-\frac{i}{2\,\tau}\,\pm\,\sqrt{c^2k^2-\frac{1}{4\tau^2}}
\label{omega}
\end{equation}
which is shown in Fig. \ref{fig0}.

An important property is the emergence of the gap in $k$-space: in order for $\omega$ in (\ref{omega}) to be real, $k>k_g$ should hold, where the $k$-gap is
\begin{equation}
k_g=\frac{1}{2c\tau}
\label{kgap}
\end{equation}

\noindent increases with the temperature because $\tau$ decreases.

The gap in $k$-space, or momentum space is interesting. Indeed, the two commonly discussed types of dispersion relations are either gapless as for photons and phonons, $E=k$ ($c=1$), or have the energy (frequency) gap for massive particles, $E=\sqrt{k^2+\mathfrak{m}^2}$, where the gap is along the Y-axis. On the other hand, (\ref{omega}) implies that the gap is in {\it momentum} space and along the X-axis, similar to the hypothesized tachyon particles with imaginary mass \cite{tachyons}.

Recently \cite{prl}, detailed evidence for the $k$-gap in liquids and supercritical fluids was presented on the basis of molecular dynamics simulations. It has been ascertained that $k_g$ increases with the inverse of liquid relaxation time in a wide range of temperature and pressure for different liquids and supercritical fluids, as (\ref{kgap}) predicts.

Interestingly, direct experimental evidence for the $k$-gap has come not from liquids but from strongly-coupled plasma \cite{pg-exp} where large particle separations and slow time scales enabled direct imaging of plasma particles. In liquids, there are two pieces of indirect evidence for the $k$-gap. The first piece of evidence comes from the fast sound or positive sound dispersion (PSD), the increase of the measured speed of sound over its hydrodynamic value \cite{ropp}. As first noted
by Frenkel \cite{frenkel}, a non-zero shear modulus of liquids implies that the propagation velocity crosses over from its hydrodynamic value $v=\sqrt{\frac{B}{\rho}}$ to the solid-like elastic value $v=\sqrt{\frac{B+\frac{4}{3}G}{\rho}}$, where $B$ and $G$ are bulk and shear moduli, respectively \cite{landelast,dyre}. According to the discussion in the previous section, shear modes become propagating $k>k_g$, implying PSD at these $k$-points. This further implies that PSD should disappear with temperature starting from small $k$ because the $k$-gap increases with temperature. This is confirmed experimentally \cite{morkel}: inelastic X-ray experiments in liquid Na show that PSD is present in a wide range of $k$ at low temperature. As temperature increases, PSD disappears starting from small $k$, in agreement with the $k$-gap picture. At high temperature, PSD is present at large $k$ only.

Another piece of evidence comes from low-frequency shear elasticity of liquids at small scale \cite{noirez1,noirez2}. According to (\ref{omega}), the frequency at which a liquid supports shear stress can be arbitrarily small provided $k$ is close to $k_g$. This implies that small systems are able to support shear stress at low frequency. This has been ascertained experimentally \cite{noirez1,noirez2}. This important result was to some extent surprising, given the widely held view that liquids were thought to be able to support shear stress at high frequency only \cite{frenkel,dyre}.

Equations \eqref{quad}-\eqref{kgap} represent important results of this section and will be discussed in later sections related to holographic models.

\section{Holographic models}\label{sec2}

\subsection{A global symmetry framework}

In this section, we describe two holographic models which display remarkable similarities to the $k$-gap discussed in the previous section.

The first holographic model has been recently proposed in Ref. \cite{Grozdanov:2018ewh}. The setup represents the gravity dual for a finite number of elastic defects immersed in a fluid background phase and is based on recent ideas related to the role of global symmetries in electromagnetism (EM), magnetohydrodynamics (MHD) and lattice dynamics \cite{Gaiotto:2014kfa,Grozdanov:2016tdf,Grozdanov:2017kyl,Hofman:2017vwr}. The momentum is a conserved quantity, and the theory, to be renormalizable, necessitates a finite UV cutoff which will be denoted as $\mathcal{C}$.

The gravitational bulk action is defined in four dimensions as:
\begin{equation}
S\,=\,\frac{1}{2\,\kappa_4^2}\int\,d^4x\,\sqrt{g}\,\left(R\,+\,\frac{6}{L^2}\,-\,\frac{1}{12}\,H^I_{abc}H_I^{abc}\right)
\end{equation}
where $\kappa_4$ is the four-dimensional gravitational coupling and $H^I=dB^I$ the field strength of a collection of two-bulk forms labeled by the internal index $I$.

The two forms admit the simple solution in terms of their field strengths:
\begin{equation}
H^1_{txr}\,=\,H^2_{tyr}\,=\,M
\end{equation}
where $M$ physically encodes the density of the aforementioned elastic defects and all associated elastic properties of the system. The system admits a black brane solution:
\begin{align}
&ds^2\,=\,\frac{dr^2}{r^2\,f(r)}\,+\,r^2\,\left(-f(r)\,dt^2\,+\,dx^2+dy^2\right)\\
&f(r)\,=\,1\,-\,\frac{2\,M^2}{r^2}\,-\,\left(1\,-\,\frac{M^2}{2\,r_h^2}\right)\,\frac{r_h^3}{r^3}
\end{align}
where $r_h$ is the location of the black hole event horizon defined by $f(r_h)=0$.

The corresponding temperature of the dual field theory can be found as usual via the surface gravity at the horizon and reads:
\begin{equation}
T\,=\,\frac{r_h}{4\,\pi}\,\left(3\,-\,\frac{M^2}{2\,r_h^2}\right)
\end{equation}

In summary, the system can be described in terms of three parameters: the temperature $T$, the UV cutoff $\mathcal{C}$ and the defects density $M$.

For simplicity, we will focus on the simple limit $M=0$ which corresponds zero density of elastic defects. Note that this limit will make the static shear modulus $G$ zero but the instantaneous elastic modulus is finite as in liquids\footnote{To be more precise, we define the static elastic modulus as:
\begin{equation}
G_0\,=\,Re\left[\mathcal{G}^R_{T_{xy}T_{xy}}\right]\,\left(\omega=k=0\right)
\end{equation}
and the instantaneous elastic modulus as:
\begin{equation}
G_\infty\,=\,Re\left[\mathcal{G}^R_{T_{xy}T_{xy}}\right]\,\left(\omega=\infty,k=0\right)
\end{equation}
where $\mathcal{G}^R_{T_{xy}T_{xy}}$ is the retarded Green function of the shear stress tensor. Note that in liquids $G_0=0$ while in solids it is finite. See, for example, \cite{Alberte:2016xja} for this distinction in similar holographic models.}.

Our interest lies in the correlator of the two-form current $\langle\mathcal{J}_{ij}\mathcal{J}_{kl}\rangle$ which is dual to the bulk two forms $B$. The vibrational degrees of freedom of the systems are encoded in this correlator and can be thought of as the vibration of the line defects coupled to the bulk to form $B_{\mu\nu}$.

In the hydrodynamic limit $\omega/T,k/T \ll 1$ and taking the UV cutoff to be large compared to the temperature $\mathcal{C}/T \gg 1$, the equation governing the dynamics of the vibrational degrees of freedom can be written analytically as:
\begin{equation}
\omega\left(1\,-\,\frac{\omega}{\omega_g}\right)\,+\,i\,\left(\frac{\bar{\mathcal{C}}\,-\,1}{r_h}\right)\,k^2\,=\,0\label{Jmode}
\end{equation}
where $\bar{\mathcal{C}}\equiv \mathcal{C}/r_h$ is the dimensionless renormalization scale and:
\begin{equation}
\omega_g\,=\,\frac{r_h}{\bar{\mathcal{C}}\,-\,1\,+\,\frac{1}{2}\,\left(\log 3\,-\,\frac{\pi}{3\,\sqrt{3}}\right)}
\end{equation}

Eq. \eqref{Jmode} can be solved explicitly, giving:
\begin{equation}
\omega_{\pm}\,=\,-\,\frac{\omega_g}{2}\,i\,\pm \sqrt{\frac{(\bar{C}-1)\,\omega_g}{r_h}\,k^2\,-\,\frac{\omega_g^2}{4}}\label{disp}
\end{equation}
which is identical to the solution of the Maxwell-Frenkel equation in liquids
\begin{equation}
\omega\,=\,-\,\frac{i}{2\,\tau}\,\pm\,\sqrt{c^2\,k^2\,-\,\frac{1}{4\,\tau^2}}
\end{equation}
with a finite $k$-gap \eqref{omega}. Moreover, we immediately obtain the expressions for the speed $c$ and relaxation time $\tau$:
\begin{equation}
c^2\,=\,\frac{(\bar{C}-1)\omega_g}{r_h}=\frac{9 (4 \pi -3 \tilde{C})}{2 \pi  \left(18+\sqrt{3} \pi -9 \log (3)\right)-27 \tilde{C}}\label{fast}
\end{equation}
\begin{equation}
\tau\,\equiv \frac{1}{\omega_g}\,=\,\frac{27 \,\tilde{C}\,-\,2\, \pi  \left(18+\sqrt{3} \pi -9 \log (3)\right)}{48 \,\pi ^2\, T}\label{tree}
\end{equation}
where we have used $\tilde{C}\equiv \mathcal{C}/T$.

Let us highlight two important features of the model: (a) at infinite cutoff $\tilde{C}\rightarrow \infty$, the speed becomes relativistic $c=1$, the relaxation time $\tau$ diverges and the $k$-gap closes as a result of (\ref{disp}) and (b) the $k$-gap increases with temperature and the relaxation time decreases as $T^{-1}$. This behavior appears similar to what takes place in liquids.

Using standard methods, we can see that the Green function of the conserved two form current $J_{\mu\nu}$ contains a single hydrodynamic mode:
\begin{equation}
\omega\,=\,-\,i\,D\,k^2\,+\,\dots\label{diff}
\end{equation}
which displays a diffusive dispersion relation.

The diffusion constant is given by:
\begin{equation}
D\,=\,\frac{3}{4\,\pi\,T}\,\left(\frac{3\,\mathcal{C}}{4\,\pi\,T}\,-\,1\right)
\end{equation}
It is straightforward to check that the following relation holds
\begin{equation}
\tau\,=\,\frac{D}{c^2}
\end{equation}
The latter can be easily derived expanding the $k$-gap dispersion relation \ref{omega} at low momenta and matching it with the diffusive mode \ref{diff}. We propose that this is a universal relation related to the appearance of the $k$-gap in relativistic systems which could have potentially important consequences.

Let us emphasize one final feature of the model. At zero density $M=0$, the shear sector governed by the stress tensor and the dynamics of the two-form current are decoupled. As a consequence, the relaxation time appearing in Eq. \eqref{tree} cannot be determined by viscosity and elasticity of the system. In other words, the relaxation time appearing in the $k$-gap dispersion relation in this model does not need to be identified with Maxwell relaxation time \eqref{tmax}. This suggests that the physical behaviour related to Eq. \eqref{quad} is more general than the Maxwell viscoelastic model and is potentially applicable to different physical systems.

\begin{figure}[h]
{\scalebox{0.45}{\includegraphics{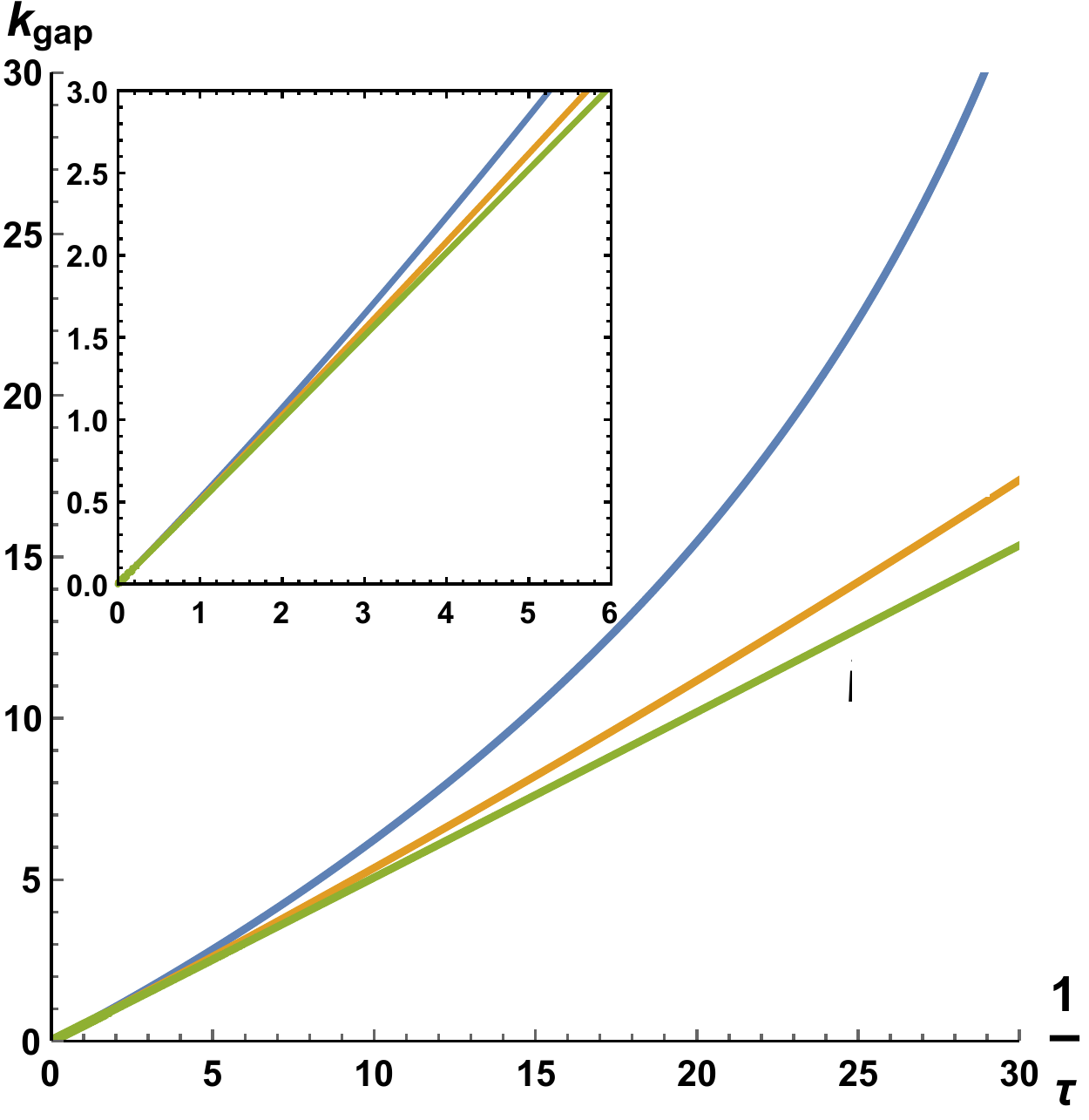}}}
\caption{The $k$-gap as a function of $\frac{1}{\tau}$ for the holographic model of Ref. \cite{Grozdanov:2018ewh}. Different colors are for different cutoffs $\mathcal{C}/T=10,50,1000$ (blue,orange,green). The straight dashed line is a guide for the eye.}
\label{fig3b}
\end{figure}

We can finally ascertain the validity of the main result for the $k$-gap (\ref{kgap}) predicting $k_g\propto\frac{1}{\tau}$. We plot $k_g$ as a function of $\frac{1}{\tau}$ in Figure \ref{fig3b}. We observe that the dependence is linear for large cutoff. For smaller cutoff, $k_g$ vs $\frac{1}{\tau}$ departs from linearity. The analogy with liquids can explain this departure as follows. As discussed earlier, smaller cutoff in the HM corresponds to smaller activation barrier in liquids and hence smaller $\tau$, resulting in faster-than-linear increase of $k_g$ according to (\ref{kgap}).

\subsection{The linear axions theory and its collective modes}
We now focus on a second holographic model known as \textit{the linear axions model} \cite{Andrade:2013gsa}. This represents a simple setup to introduce momentum relaxation into the holographic framework using a specific massive gravity theory in the bulk \cite{Vegh:2013sk,Baggioli:2014roa}. The four dimensional gravitational bulk action is defined as:
\begin{equation}
S\,=\, M_P^2\int d^4x \sqrt{-g}
\left[\frac{R}2+\frac{3}{\ell^2}- \, m^2 \partial_\mu \phi^I \partial^\mu \phi^I\right]
\end{equation}
where $M_P$ is the Planck mass and the $\phi^I$ are two St\"uckelberg fields \cite{Alberte:2015isw}. $\phi^I$ have a radially constant bulk profile $\phi^I=x^I$ with $I=x,y$.
This is an exact solution due to the shift symmetry of the system. In the dual description, these fields are marginal operators breaking translational invariance because of their explicit dependence on the spatial coordinates. The geometry remains homogeneous thanks to the global symmetries of the setup and can be written in the simple form:
\begin{equation}
\label{backg}
ds^2=\frac{\ell^2}{u^2} \left[\frac{du^2}{f(u)} -f(u)\,dt^2 + dx^2+dy^2\right] ~,
\end{equation}
where $u\in [0,u_h]$ is the radial holographic direction spanning from the boundary to the horizon, defined through $f(u_h)=0$, and $\ell$ is the AdS radius.

The blackening factor appearing in the black brane solution above takes the form:
\begin{equation}
f(u)\,=\,1\,-\,m^2\,u^2\,\,-\,\frac{u^3}{u_h^3}\,+\,m^2\,\frac{u^3}{u_h}
\end{equation}
and the corresponding temperature reads:
\begin{equation}
T\,=\,\frac{3}{4\,\pi\,u_h}\,-\,\frac{m^2\,u_h}{4\,\pi}
\end{equation}

The dual theory of the bulk model \cite{Andrade:2013gsa} displays a finite relaxation time for the momentum operator $\tau_{rel}^{-1}=\Gamma$ which is inversely proportional to the graviton mass $\sim m^2$ (\cite{Davison:2013jba}). From the symmetry point of view, this can easily be understood: the graviton mass breaks explicitly diffeomorphisms invariance of the bulk theory, and the conservation of the dual stress tensor does not hold as a result. It is important to note that the stress tensor does not acquire an anomalous dimension because the graviton mass is zero at the UV fixed point.

\begin{figure}[h]
{\scalebox{0.4}{\includegraphics{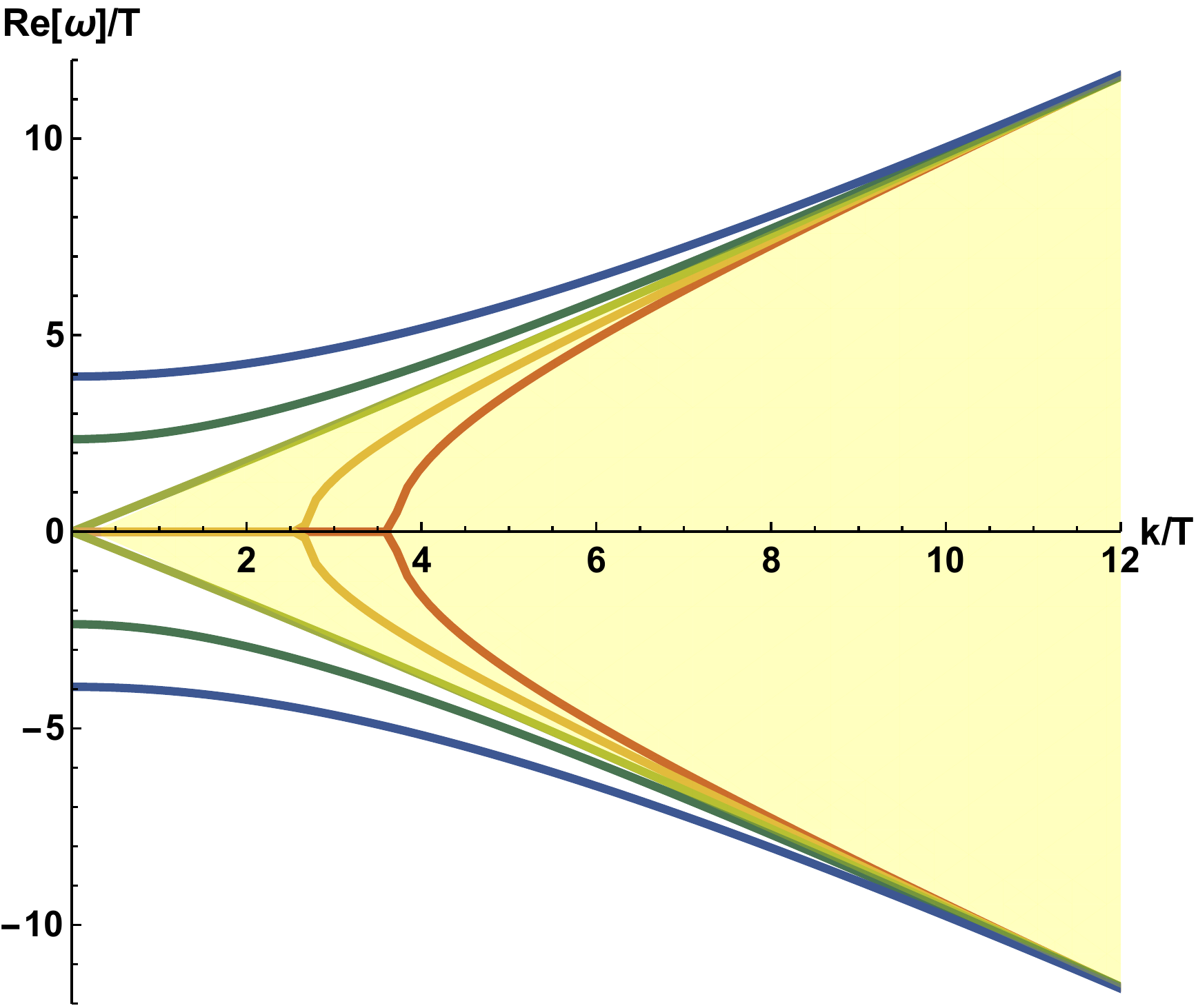}}}
\caption{The dispersion relation of the transverse collective modes obtained numerically in the HM of \cite{Andrade:2013gsa}. The temperature increases from blue ($T/m \sim 0.141$) to red ($T/m \sim 0.171$). The shaded region ($T/m>0.156$) displays the presence of the $k$-gap.}
\label{fig1}
\end{figure}

The computation of the stress tensor correlator $\langle T_{xy}T_{xy}\rangle$ can be done by solving the equations for perturbations on the background shown previously. The transverse or shear perturbations are encoded in the fluctuations $a_x,\,h_{tx}\equiv u^2 \delta g_{tx},\,h_{xy}\equiv u^2\delta g_{xy},\,\delta \phi_x,\,\delta g_{xu}$. Assuming the radial gauge, \textit{i.e.} $\delta g_{xu}=0$, and using the ingoing Eddington-Finkelstein coordinates
\begin{equation}
ds^2=\frac{1}{u^2} \left[-f(u)dt^2-2\,dt\,du + dx^2+dy^2\right]
\end{equation}
the remaining equations read
\begin{align}
&-2h_{tx}+u\,h_{tx}'-i\,k\,u\, h_{xy}-\left( k^2\,u+2\,i\,\omega\right)\,\delta \phi_x\nonumber\\&+u\,f\, \delta \phi_x''+\left(-2\,f+u\,(2 i \omega+f')\right)\delta \phi_x'\,=\,0
\,\,;\nonumber\\[0.1cm]
&u^2\, k\,\omega\, h_{xy}+\left(6+k^2\,u^2 \, -6 f+2uf'\right) \, h_{tx}
\nonumber\\&+2\,i\,m^2\,u^{2}\omega\,\delta \phi_x+\left(2\,u\,f-i\,u^2\omega\right)h_{tx}'-u^2\,f\,h_{tx}''\,=\,0 \,\,;
\nonumber\\[0.1cm]
&-iku^2h_{tx}'-2\,i \, k \,m^2 \,u^{2}\delta \phi_x+2 h_{xy}\left(3+i\,u \, \omega-3f+uf' \right)
\nonumber\\&+2i\,k\,u\,h_{tx}-\left(2i\,u^2 \, \omega-2uf+u^2\,f'\right)h_{xy}'-u^2\,f\, h_{xy}''\,=\,0\,
\nonumber\\[0.1cm]
&2\,h_{tx}'-u\,h_{tx}''-2m^2\,u \, \delta\phi_x'+ik\,u\,h_{xy}'\,=\,0\,\,.
\end{align}
and their solution can be obtained numerically.

Importantly, the asymptotic behaviour of the different bulk fields close to the UV boundary $u=0$ are:
\begin{align}
&\delta \phi_x\,=\,\phi_{x\,(l)}\,(1\,+\,\dots)+\,\phi_{x\,(s)}\,u^{3}\,(1\,+\,\dots)\,,\quad\nonumber\\ &h_{tx}\,=h_{tx\,(l)}\,(1\,+\,\dots)\,+\,h_{tx\,(s)}\,u^{3}\,(1\,+\,\dots)\,,\nonumber\\ &h_{xy}\,=h_{xy\,(l)}\,(1\,+\,\dots)\,+\,h_{xy\,(s)}\,u^{3}\,(1\,+\,\dots)\,.
\end{align}
where the subscript $l$ stands for ''leading'' and the subscript $s$ for ''subleading'' contributions.
Choosing this coordinates system, the ingoing boundary conditions at the horizon are automatically satisfied by regularity at the horizon.
The various retarded Green's functions can be defined as:
\begin{align}\label{greenF}
&\mathcal{G}^{\textrm{(R)}}_{T_{tx}T_{tx}}\,=\,\frac{2\,\Delta-d}{2}\,\frac{h_{tx\,(s)}}{h_{tx\,(l)}}\,=\,\frac{3}{2}\frac{h_{tx\,(s)}}{h_{tx\,(l)}}\,,\nonumber\\
&\mathcal{G}^{\textrm{(R)}}_{T_{xy}T_{xy}}\,=\,\frac{2\,\Delta-d}{2}\,\frac{h_{xy\,(s)}}{h_{xy\,(l)}}\,=\,\frac{3}{2}\frac{h_{xy\,(s)}}{h_{xy\,(l)}}\,.
\end{align}
where spacetime dependences are omitted for simplicity. The conformal dimension of the stress tensor operator is simply $\Delta=3$. From the poles of the Green functions, defined as the zero of the leading terms in the UV expansions, we can read off the QNMs frequency at a finite momentum.

The first important result from the first model is the emergence of the $k$-gap shown in fig \ref{fig1}. The numerical study of the transverse fluctuations results in the identification of the QNMs spectrum shown in Fig. \ref{fig1}. At high enough temperature, we observe the emergence of the gap in $k$-space. The $k$-gap increases with temperature, the same effect as in liquids (see (\ref{kgap})) derived from the Maxwell-Frenkel approach.

The temperature at which the $k$-gap opens up in Fig. \ref{fig1} corresponds to $T/m>0.156$. At smaller values of $T/m$, the spectrum has a mass gap as is the case for a massive particle. This can be attributed to the competition between the effective mass term  $\mathfrak{m}$ and the dissipative $\frac{1}{\tau}$ term: adding the mass term to the Lagrangian describing the $k$-gap in (\ref{omega}) modifies the dispersion relation as \cite{pre}:

\begin{equation}
\omega=\sqrt{k^2+\mathfrak{m}^2-\frac{1}{4\tau^2}}
\label{km}
\end{equation}

According to (\ref{km}), the dispersion relation is linear and gapless for $\mathfrak{m}=\frac{1}{2\tau}$, whereas the gap in $k$-space opens up for $\frac{1}{2\tau}>\mathfrak{m}$. The nature of the effective mass $\mathfrak{m}$ within the model \cite{Andrade:2013gsa} will be discussed elsewhere.

At a specific value of $T/m$, at which the energy density vanishes $\epsilon=0$, the system enjoys an enhanced symmetry which allows us to compute the Green function for the transverse modes analytically \cite{Davison:2014lua}. As a result we are able to find analytically the $k$-gap as:
\begin{equation}
\omega\,=\,-\frac{3}{2}\,i\,\pm\,\sqrt{k^2\,-\,\frac{1}{4}}\label{resres}
\end{equation}

\begin{figure}[h]
{\scalebox{0.58}{\includegraphics{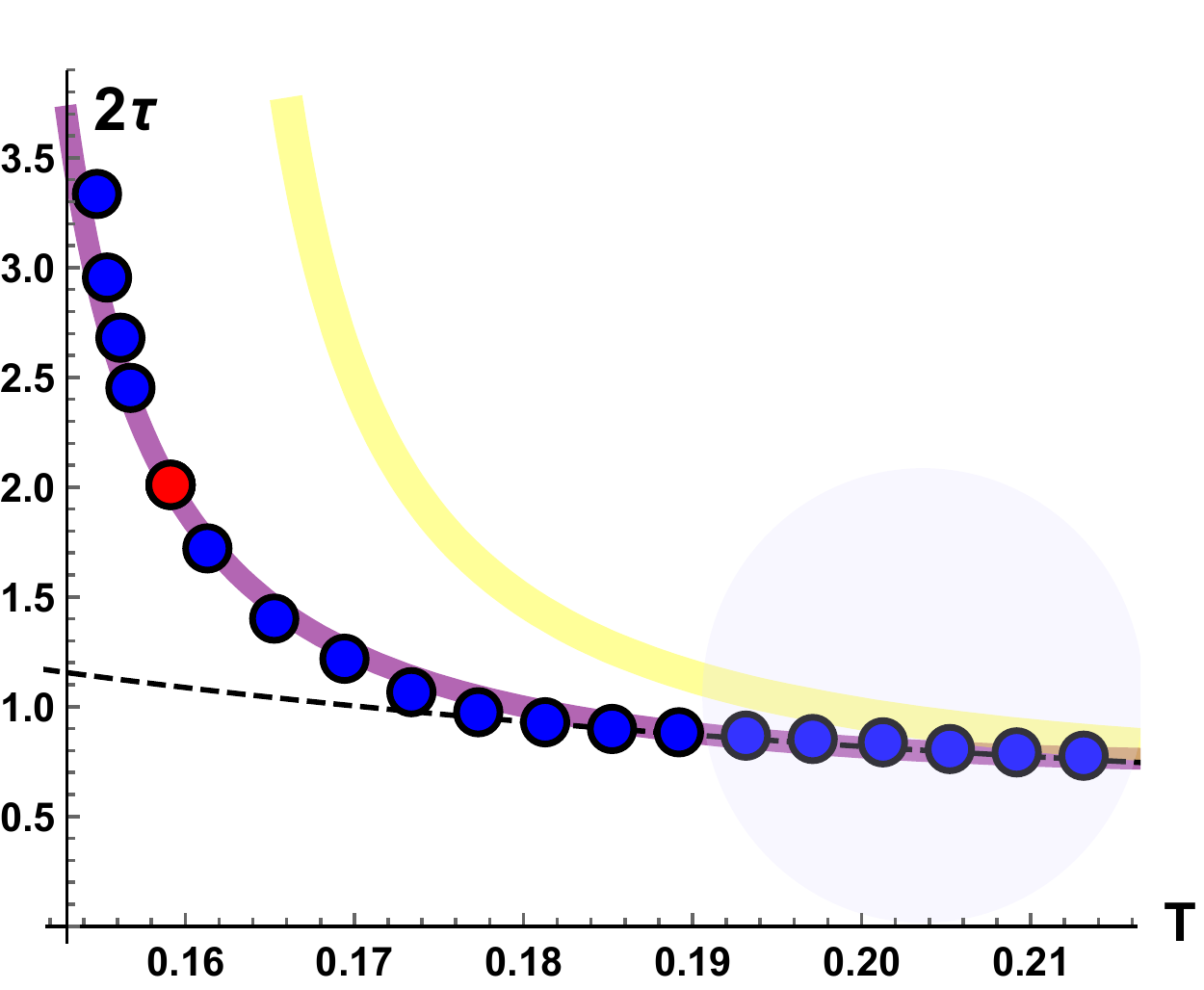}}}
\caption{Comparison between the numerical data (filled bullets) obtained fitting the dispersion curves $\omega(k)$ and the analytic formula \eqref{gen} (purple line) for the HM of \cite{Andrade:2013gsa}. We also show the value of the naive timescale $D/c^2$ (yellow line) and the approximate formula of \cite{Ciobanu:2017fef} (dashed line). We plot just an intermediate scale of temperatures because away from it our formulas are not valid. At low $T/m$ momentum dissipation is strong and the shear mode becomes overdamped. At high $T/m$ the relaxation time $\tau$ becomes small and therefore the second pole becomes overdamped. The red dot is the result at the self-dual point at which the k-gap can be extracted analytically \eqref{resres}.}
\label{fig2}
\end{figure}

The corresponding $\tau$, calculated by matching the $k$-gap dispersion relation \eqref{omega} with the analytic poles of the Green function (see \eqref{resres}), is shown in red in Fig. \ref{fig2} and agrees well with $\tau$ calculated from (\ref{gen}). Finally, we note that in the limit of large $T/m$, analytic formulas for the diffusion constant $D$ and the momentum dissipation rate $\Gamma$ have been obtained \cite{Davison:2013jba,Ciobanu:2017fef}:
\begin{align}
&\Gamma\,=\,\frac{m^2}{2\,\pi\,T}\,+\,\dots\\
&D=\frac{1}{4\pi T}\left[1+\frac{1}{24}\,\left(9+\sqrt{3}\pi-9\log 3\right)\frac{m^2}{8\pi^2\,T^2}\right]+\dots
\end{align}

Both previous formulas are in agreement with our numerical results in the regime $T/m \gg 1$.

Before proceeding with the analysis of the $k$-gap further, we discuss the relaxation time $\tau$ and its physical meaning in liquids and HMs. There are two dynamical regimes in the liquid state where liquid properties are qualitatively different. In the low-temperature regime, particle dynamics combines solid-like oscillatory motion and diffusive jumps between quasi-equilibrium positions. In the low-temperature regime, the diffusion constant $D$ is approximately $D_l=\frac{a^2}{\tau}$, where $a$ is of the order of interatomic separations, and decreases with $\tau$. In this regime, $D_l$ is inversely proportional to viscosity $\eta$, and $\eta$ itself decreases with temperature \cite{frenkel}. In the high-temperature gas-like regime, particles lose the oscillatory component of their motion and move as in a gas. In this regime, the relaxation time $\tau$ is related to the time between momentum-transferring particle collisions which sets the length of the mean free path. The diffusion constant $D_h=\frac{\eta}{\rho}$ is proportional to $\eta$, and $\eta$ itself increases with temperature \cite{frenkel}. The liquid-like molecular motion combining solid-like oscillatory and gas-like diffusive motion is separated from purely diffusive molecular motion by the Frenkel line recently introduced and extending to the supercritical state of matter \cite{ropp}.

A hydrodynamic description of relativistic fluids is characterised by (a) proportionality between $D$ and $\eta$ as $D=\eta/(\epsilon+p)$ and (b) viscosity increasing with temperature. As discussed above, this implies that these systems are in the high-temperature gas-like dynamical regime from the point of view of condensed matter physics. The relationship between $D$ and $\tau$ in this regime can be derived by expanding $\omega$ in \eqref{omega} in the hydrodynamic limit of small $k$ $\omega_-\,=\,-\,i\,c^2\,\tau\,k^2\,+\,\dots$ or solving Eq. (\ref{gener3}) in its Navier-Stokes form without the second time derivative term and subsequently comparing the result with the usual diffusive mode $\omega=-i D k^2$. This gives a simple relation\footnote{The same expression was already considered in \cite{Pu:2009fj} in the study of the causality of relativistic dissipative fluid dynamics.}
\begin{equation}
D\,=\,c^2\,\tau \label{Dformula}
\end{equation}

We note that relaxation time $\tau$ can also be obtained from $Im(\omega)$ at large $k$, where $\omega$ is given by the dispersion relation \eqref{omega}. $Im(\omega)$ approaches a constant value in the limit of large $k$ (see \cite{Baggioli:2018vfc} for more details).

The same relation (\ref{Dformula}) follows from considering $\tau$ in $D_l=\frac{a^2}{\tau}$ as the time between particle collisions in the high-temperature regime, in which case $a=c\tau$ becomes the distance travelled ballistically. For relativistic fluids, (\ref{Dformula}) follows from combining $D=\eta/(\epsilon+p)$ with the speed of transverse phonons $c^2=G/(\epsilon+p)$ to yield $D=c^2\frac{\eta}{G}$ and subsequently noting that $\frac{\eta}{G}$ is relaxation time from Maxwell theory\footnote{We note that in the Maxwell viscoelastic model, the shear modulus $G$ corresponds to the instantaneous modulus $G_{\infty}$ and therefore we do not need to distinguish them with different symbols.}.

The last point calls for two important observations related to Maxwell interpolation (\ref{a1}) which gives rise to the $k$-gap equation \eqref{quad}. Notably, Maxwell interpolation was discussed and later developed in the low-temperature liquid-like dynamical regime only, with $G$ being the solid-like high-frequency shear modulus governed by interatomic interactions. However, we propose that Maxwell interpolation also applies to the high-temperature gas-like regime. Indeed, the idea of the system being able to support two types of response, viscous and elastic, is general enough and applicable to the high-temperature gas-like state as well, but with the proviso that in this state $G$ in (\ref{a1}) describes a purely kinetic term $\propto T$ due to particle inertia ($\eta$ is defined in the usual way).

Second, the emergence of the $k$-gap in liquids due to Maxwell interpolation differs from the HMs in one important respect. In liquids as well as solids, there is an ``ultraviolet'' cutoff related to the shortest interatomic separation $a$ and the largest, Debye, frequency in the system, $\omega_{\rm D}$, or the shortest vibration period, $\tau_{\rm D}$. When $\tau$ in (\ref{kgap}) reduces at high temperature and approaches $\tau_{\rm D}$, the $k$-gap extends to the entire first zone or, equivalently, the wavelength of propagating shear modes becomes comparable to the interatomic separation. At this point, all transverse modes disappear from the liquid spectrum \cite{ropp}. On the other hands, no equivalent cutoff exists in HMs. Therefore, the $k$-gap for propagating shear modes in HMs is not bounded from above.

We now return to our analysis of properties of the $k$-gap. The model has an additional dissipative contribution to $\tau$ due to momentum relaxation time $\tau_{rel}=\Gamma^{-1}$, resulting in $\omega=-i \Gamma-i D k^2$. As a consequence, the description in terms of Eq. \eqref{quad} becomes more subtle. First, the requirement for the validity of Eq. \eqref{quad} is that relaxation time $\tau$ is large compared to the characteristic energy scale of the system $\tau T \gg 1$. The latter guarantees that the mode $\omega=- i \tau^{-1}$ is underdamped and is included in the hydrodynamic description. In the opposite scenario when $\tau T \ll 1$, such a mode is over-damped and is not present in the hydrodynamic description. The second requirements is that the shear diffusive mode is not over-damped because of momentum relaxation. This implies that the momentum relaxation rate is small, \textit{i.e.} $\tau_{rel}T \gg 1$.

In order for the two above conditions to hold, we find that Eq. \eqref{quad} can accurately describe physics of the system only in an intermediate regime of $m/T$, as is shown in Fig. \ref{fig2}. Interestingly, we find that at lower values of $T/m$ where momentum relaxation mechanism becomes important, the relaxation time $\tau$ formula is in good agreement with

\begin{equation}
\tau\,=\,\frac{D}{c^2\,-\,D\,\Gamma}\label{gen}
\end{equation}
which can be obtained by applying a simple inverse Matthiesen'rule. Finally, we note that momentum relaxation is not necessary for the emergence of the $k$-gap.

We now discuss the numerical results of this model in more detail. We fit the calculated dispersion curves with $k$-gaps in Fig. \ref{fig1} to $\omega$ predicted on the basis of Maxwell interpolation in Eq. (\ref{omega}), fixing the speed to its relativistic limit $c=1$. We find that the fits are of high quality. Using the fits, we extract the corresponding $\tau$ in (\ref{omega}) and plot $\tau$ in Figure \ref{fig2}. We subsequently compute $D$ from the imaginary part of $\omega$ at low momenta numerically and calculate $\tau$ using (\ref{Dformula}) and (\ref{gen}) as well as an approximate analytical equation for $D$ \cite{Ciobanu:2017fef}. The resulting curves are shown in Figure \ref{fig2}. We observe that $\tau$ from the fit of $\omega$ to (\ref{omega}) and $\tau$ calculated from Eq. (\ref{gen}) agree with high accuracy. We further observe that relaxation time $\tau$ obtained by fitting the numerical data and all $\tau$ calculated from the diffusion constant $D$ coincide in the intermediate temperature regime already mentioned and displayed with a shaded region. At lower temperature, we note that Eq. \eqref{gen} captures the behaviour of the numerical data, representing a good approximation that takes the first corrections due to the momentum relaxation rate $\Gamma \neq 0$ into account.

Several important implications follow from our analysis and from Figure \ref{fig2}. First, $\tau$ decreases with $T$ as is the case in liquids. Second, the dispersion relations in the HMs with the $k$-gap emerging in Fig. \ref{fig1} agree with the liquid gapped dispersion relation resulting from Maxwell interpolation (\ref{omega}) as discussed above. This, in turn, suggests that the ideas involved in Maxwell interpolation and its extension by Frenkel can be used more generally to analytically treat HMs and strongly-coupled fields.

\begin{figure}[h]
{\scalebox{0.45}{\includegraphics{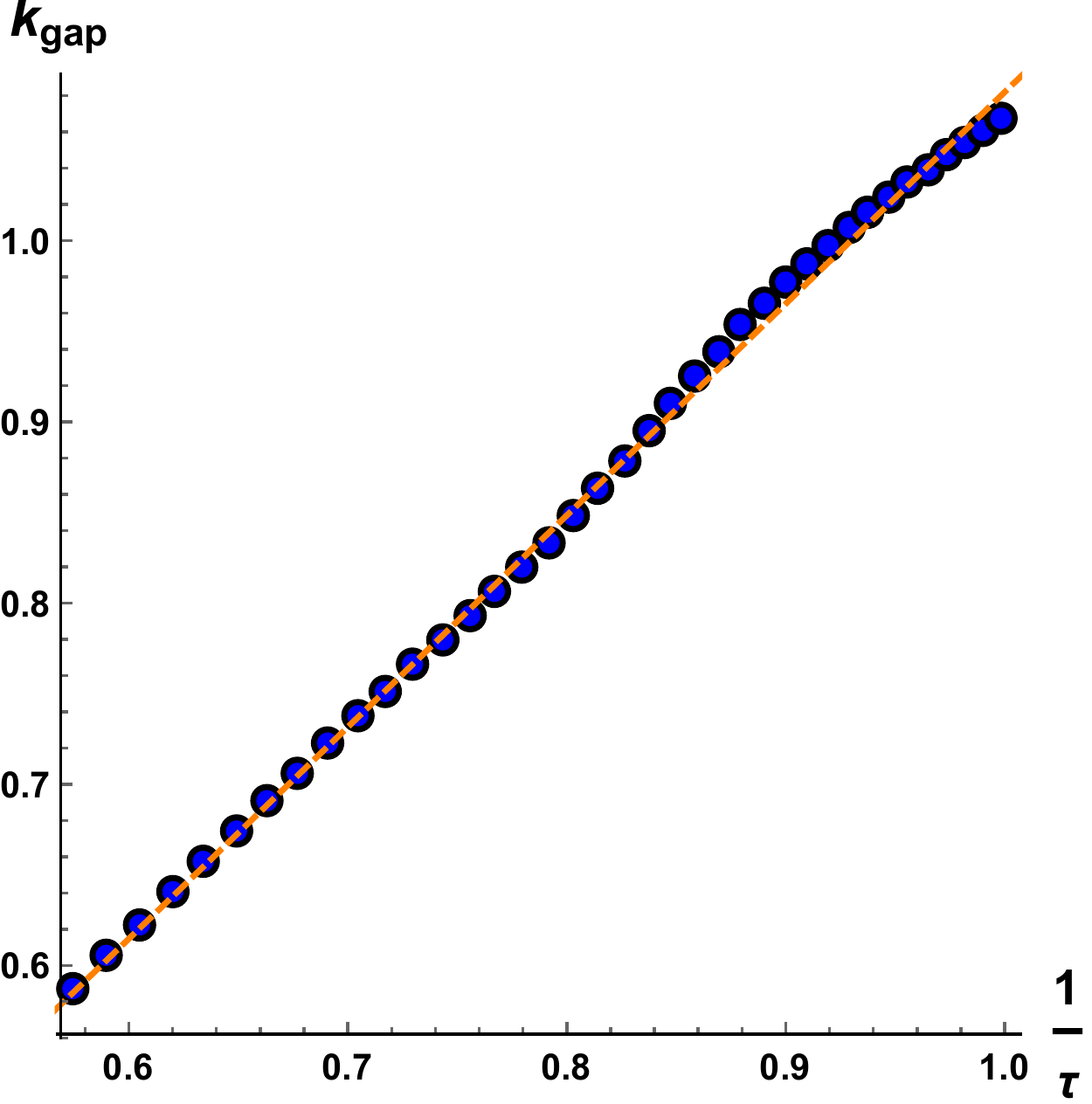}}\vspace{0.5cm}}
\caption{The $k$-gap as a function of $\frac{1}{\tau}$ in the HM of \cite{Andrade:2013gsa}.}
\label{fig3c}
\end{figure}

Finally, we observe the validity of the main result for the $k$-gap (\ref{kgap}) predicting $k_g\propto\frac{1}{\tau}$. We plot $k_g$ as a function of $\frac{1}{\tau}$ in Figure \ref{fig3c} and observe a linear dependence with good accuracy.

We make a remark regarding relaxation time $\tau$ that governs the $k$-gap in the Maxwell-Frenkel approach. From a practical perspective, the introduction of $\tau$ in (\ref{a1})-(\ref{kgap}) enables us to discuss collective modes in liquids and their role in liquid dynamical and thermodynamic properties \cite{ropp}. From a general-theoretical perspective of treating strongly-interacting and dynamically disordered systems, the introduction of $\tau$ simplifies and solves an exponentially complex problem of coupled non-linear oscillators describing the motion of liquid particles in the strongly anharmonic multi-well potential \cite{ropp}. Although not derived from first-principles (recall that a first-principles description of liquids is exponentially complex and hence is non-tractable), $\tau$ is an important liquid property directly linked to viscosity that enables to provide relationships between different liquid properties \cite{ropp}. Therefore, the introduction of $\tau$ is a non-perturbative way to treat strong interactions and in this sense is particularly suitable to address strong interactions in field theories.

Before concluding, we note that the $k$-gap has earlier appeared in the Israel-Stewart formalism for relativistic hydrodynamics \cite{ISRAEL1979341,Romatschke:2009im}. In these studies, the second-order terms of the gradient expansion introduce a relaxation time in order to cure the linearized approximation from causality and unitarity issues \cite{Pu:2009fj}. In our case, the relaxation time has the physical meaning and directly affects the physical degrees of freedom. In this sense, our results are different from those obtained in the Israel-Stewart framework where introducing relaxation time is a phenomenological step which does not appear in the full description of relativistic hydrodynamics and does not affect realistic collective modes.

\section{Conclusions}\label{sec3}
In summary, we have shown that liquids and holographic models are strikingly similar in terms of several detailed and specific properties. Similarly to liquids, we find that (a) the HMs develop the $k$-gap with the same dispersion relation, (b) the $k$-gap in the HMs increases with temperature, (c) the relaxation time $\tau$ governing the $k$-gap in the HMs coincides with the relaxation time calculated from the gapped dispersion relation following from Maxwell interpolation, (d) $\tau$ in the HMs decreases with temperature as in liquids, and (e) the $k$-gap is inversely proportional to the relaxation time. These close similarities suggest that the general idea involved in Maxwell interpolation and its Frenkel development can serve as a constructive approach to treat holographic models and their strongly-coupled field theory counterparts. It will be interesting to construct a more general theoretical framework to explain the $k$-gap phenomenon and its appearance in different areas of physics.

\section*{Acknowledgments}
We thank A. Zaccone, L. Noirez, Richard Davison and Ben Withers for useful discussions and comments about this work. We are particularly grateful to Martin Ammon and Amadeo Jimenez for help, support and discussions and for sharing with us the numerical codes used in \cite{Alberte:2017cch}.
This research utilised Queen Mary's Apocrita HPC facility, supported by QMUL Research-IT.
KT is grateful to the EPSRC for support.
MB is supported in part by the Advanced ERC grant SM-grav, No 669288.
MB would like to thank Marianna Siouti for the unconditional support.

\end{document}